\documentclass{pic2012}
\def\runauthor{Alessia Tricomi}
\def\shorttitle{Very Forward proton-proton interactions with the LHCf detector}
\usepackage{amsmath}
\usepackage{amssymb}
\begin{document}

\title{VERY FORWARD PROTON-PROTON INTERACTIONS WITH THE LHCF DETECTOR}

\author{Alessia Tricomi on behalf of the LHCf Collaboration}
\address{Dipartimento di Fisica e Astronomia and INFN Catania\\
Catania, Italy\\
E-mail: alessia.tricomi@ct.infn.it }
\maketitle

\abstracts{The LHCf experiment has been designed to precisely measure very forward neutral particle spectra 
produced in proton-proton collisions at LHC up to an energy of 14 TeV in the center of mass system. 
These measurements are of fundamental importance to calibrate the Monte Carlo models widely used in the 
high energy cosmic ray (HECR) field, up to an equivalent laboratory energy of the order of $10^{17}$ eV. \\
In 2009-2010 the experiment has completed the p-p data taking at $\sqrt{s} = 0.9$ TeV and $\sqrt{s} = 7$ TeV and 
the detectors have later on been removed from the tunnel region, when the LHC luminosity 
increased above $10^{30}$ cm$^{-2}$s$^{-1}$. \\
In this paper the most up-to-date results on the inclusive photon spectra and the $\pi^0$ spectra measured by LHCf are reported. 
Comparison of these spectra with the model expectations and the impact on high energy cosmic ray (HECR) Physics are discussed.
In addition, perspectives for future analyses as well as the 
program for the next data taking period, in particular the foreseen data taking in p-Pb collisions, will be discussed.}

\section{Introduction} 
LHCf is the smallest of the LHC experiments. It has been designed 
to study neutral particles produced 
in proton-proton collisions at LHC in the very forward region, covering the $|\eta|>8.3$ 
pseudo-rapidity region. 

The LHCf experiment differs from the other LHC experiment not only in dimensions but also 
for the main physics motivation, which for LHCf is strictly connected to astroparticle physics.
The goal of the experiment is indeed to measure neutral particle spectra to 
calibrate Monte Carlo codes used in High Energy Cosmic Ray (HECR) physics. A good knowledge of 
nuclear interaction model of primary cosmic rays with earth's atmosphere is mandatory to 
better understand many properties of primary cosmic rays, like the 
energy spectrum and the composition, whose knowledge is finally strictly related to our 
capability to understand the origin of high energy phenomena in the Universe.
 Dedicated extensive air shower experiments are taking data since many years and 
have strongly contributed to our understanding of High and Ultra High Energy 
Cosmic (UHECR) Ray Physics. However, the results 
of these experiments are in some cases not fully in agreement and, in addition, 
the interpretation of their data in terms of primary cosmic ray properties 
is strongly affected by the knowledge of the nuclear interactions   
in the earth's atmosphere. This is true, for instance, for the interpretation  
of the behaviour of the energy spectrum in the UHE region, in particular the 
existence of events above the so called GZK cut-off, and the chemical composition 
of cosmic rays.  Indeed, evidence of UHECR, above the GZK cut-off, has been reported
for the first time by the AGASA experiment~\cite{Takeda}.
On the contrary, the results of the HiRes~\cite{Abbasi:2007sv} experiment and, more recently, the ones of 
the Pierre Auger Collaboration~\cite{Abreu:2011pj} and the Telescope Array Collaboration~\cite{Tsunesada:2011mp} 
are consistent with the existence of the cut-off. 
The disagreement among data would be 
reduced by adjusting the energy scales of the different experiments to account for 
systematic effects in the determination of 
the particle energy, that might be due to different detecting 
techniques. Similar considerations hold for the interpretation of 
cosmic ray composition since it is directly related to their 
primary sources. Accelerator experiments validating the interaction model chosen are hence essential. 
As a matter of fact air shower development is dominated by the
forward products of the interaction between the
primary particle and the atmosphere. The only available data on the 
production cross-section of neutral pions emitted in the
very forward region have been obtained more than twenty years ago by the 
UA7 Collaboration~\cite{UA7} at the CERN Sp${\mathrm{\overline p}}$S up to an energy of 
10$^{14}$ eV and in a very narrow pseudo-rapidity range. 
The LHCf experiment at LHC has the unique opportunity to take data at 
energies ranging from $\sqrt{s} = 0.9$ TeV up to 14 TeV, thus extending significantly 
the energy range up to a region of great interest for high energy cosmic rays, the region 
between the ``knee'' and the GZK cut-off.

\section{The LHCf detector}
The detector consists of a double arm -- double tower  
sampling and imaging calorimeter, placed at $\pm$ 140 m from ATLAS interaction point (IP1) 
inside the zero-degree neutral absorbers (Target Neutral Absorber, 
TAN). Charged particles from the IP are swept away by the inner beam separation dipole 
before reaching the TAN, so that only photons mainly from $\pi^0$ decays, neutrons and 
neutral kaons reach the LHCf calorimeters. \\
Each calorimeter (ARM1 and ARM2) has a double tower structure, with the smaller tower located 
at zero degree collision angle, approximately 
covering the region with pseudo-ra\-pi\-di\-ty $\eta > 10$ and the larger one, approximately covering the region 
with $8.4 < \eta < 10$. 
Each calorimeter tower is made 
of 16 layers of plastic scintillators interleaved by tungsten layers as converter, complemented by 
a set of four X-Y position sensitive layers which provide incident shower positions, in order to 
obtain the transverse momentum of the incident primary and to 
correct for the effect of leakage from the edges of the calorimeters.  \\
The two calorimeters are identical for the calorimetric structure; however they slightly differ for the geometrical 
arrangement of the two towers and for the position sensitive layers made by 1 mm$^2$ scintillating fibers 
in one calorimeter (ARM1) and silicon micro-strip layers in the other (ARM2). \\
The two tower 
structure allows to reconstruct the $\pi^0$ decaying in two $\gamma$s, hitting separately the two towers, 
hence providing a very precise absolute energy calibration of the detectors. In the range E$> 100$ GeV, 
the LHCf detectors have energy and position resolutions for electromagnetic showers better than 5\% and $200 \mu$m, 
respectively.
A detailed description of the LHCf detector and its performance can be found in~\cite{lhcf.jinst}.

\section{Measurement of the single photon energy spectra}
\label{sec:2}

The LHCf Collaboration has measured the single photon energy spectrum in proton-proton collisions at 
7 TeV~\cite{plb} and, more recently, at 900 GeV p-p collisions~\cite{plb2}. 
Here we briefly summarize the main steps of the analysis with special emphasis to possible implication 
for the calibration of Monte Carlo models used in HECR Physics.
For both analyses only a subset of the collected 
data have been analysed corresponding to an integrated luminosity of 0.68 nb$^{-1}$ and 0.52 nb$^{-1}$ for the ARM1 and ARM2 detectors, respectively, for the 7 TeV data and to an integrated luminosity of 0.30  nb$^{-1}$ for the 900 GeV analysis.
The analysed data have been chosen in a particularly clean and low luminosity fill, to minimize backgrounds, 
hence reducing the systematics of the measurement.

The main steps of the analysis work-flow are almost identical for the two analysis. 

The energy of photons is reconstructed from the signal released by the shower particles in the scintillators,
after applying corrections for the non-uniformity of light collection and for particles 
leaking in and out of the edges of the calorimeter towers. In order to correct for these last two effects, which are 
rather important due to the limited transverse size of both the calorimetric towers, the transverse 
impact position of showers provided by the position sensitive detectors is used.  

Event produced by neutral hadrons are rejected  applying a simple particle identification algorithm based 
on the longitudinal development of the showers, which is different for electromagnetic and hadronic particles.
In addition, for the 7 TeV analysis, thanks to the information provided by the position sensitive detectors, 
events with more than one shower inside the same tower (multi-hit) are rejected, while 
for the 900 GeV analysis the number of multi-particle events is negligible hence the  multi-hit rejection is not applied.
In order to combine the spectra measured by ARM1 and ARM2, which have different geometrical 
configurations, in these analyses only events detected in a common pseudo-rapidity and azimuthal range are selected: 
$\eta > 10.94$ and $8.81 < \eta < 8.99$, for 
the small and large towers, respectively, for 7 TeV analysis and $\eta > 10.15$ and $8.77 < \eta < 9.46$, for 
the small and large towers, respectively, for 900 GeV analysis. 

\begin{figure}[hbt]
\centerline{\resizebox{125mm}{!}{\includegraphics{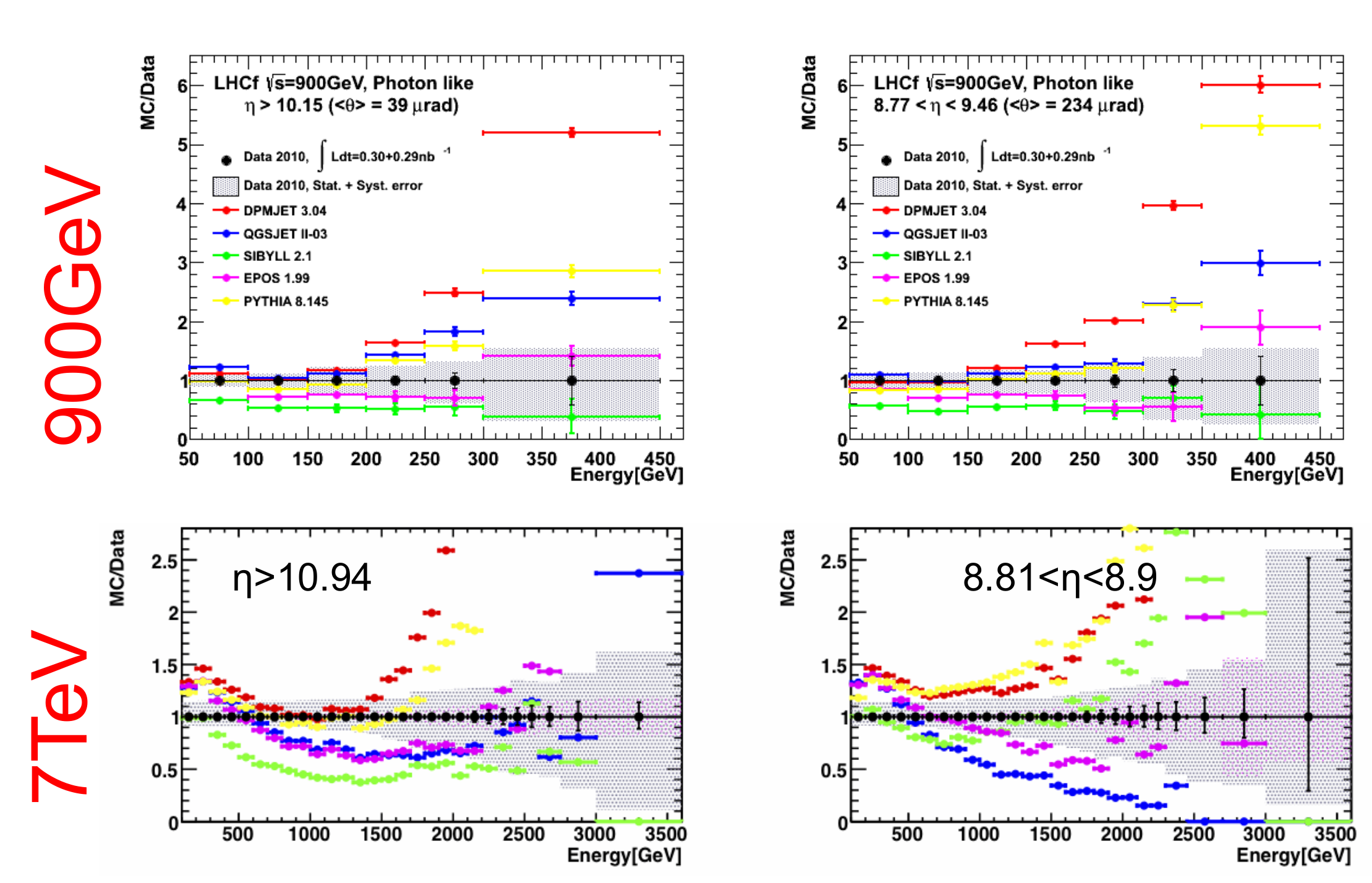}}}
\caption[*]{Single photon energy spectra measured by LHCf in $\eta > 10.94$ (left) and $8.81 < \eta < 8.99$ (right) bin, respectively, for 7 TeV (top panel) and $\eta > 10.15$ (left) and $8.77 < \eta < 9.46$ (right) bin, respectively, for 900 GeV (bottom panel) p-p collisions. The ratio of MC results predictions for DPMJET III~3.04 (red), QGSJET II-03 (blue), SIBYLL 2.1 (green), EPOS 1.99 (magenta) and PYTHIA 8.145 (yellow) to experimental data are shown in the plots. Error bars and gray shaded areas in each plot indicate the statistical and the systematic errors, respectively. Figures from Ref.~\cite{plb,plb2}.}
\label{fig.spectra}
\end{figure}

Figure~\ref{fig.spectra} shows the single $\gamma$ spectra measured by LHCf in the two pseudo-rapidity regions 
for 7 TeV and 900 GeV p-p collisions, respectively, compared with results predicted by MC simulations using different models: 
DPMJET III-3.04~\cite{DPMJET}, QGSJET II-03~\cite{qgsjet}, SIBYLL 2.1~\cite{sybill}, EPOS 1.9~\cite{epos} and 
PYTHIA 8.145~\cite{pythia}. Statical errors and systematic uncertainties are also plotted. A careful study of 
systematic uncertainties has been done and conservative estimates have been taken into account. 
Further details can be found in Ref.~\cite{plb,plb2}. As can be seen from Fig.~\ref{fig.spectra}, a clear discrepancy 
between the experimental results and the predictions of the models in particular in the high energy region is present.
The impact of such results in the tuning of MC code is now under investigation with the help of code developers.

\vfill\pagebreak
\section{Measurement of the neutral pion transverse momentum spectra}
\label{sec:3}

\begin{figure}[hbt]
\centerline{\resizebox{125mm}{!}{\includegraphics{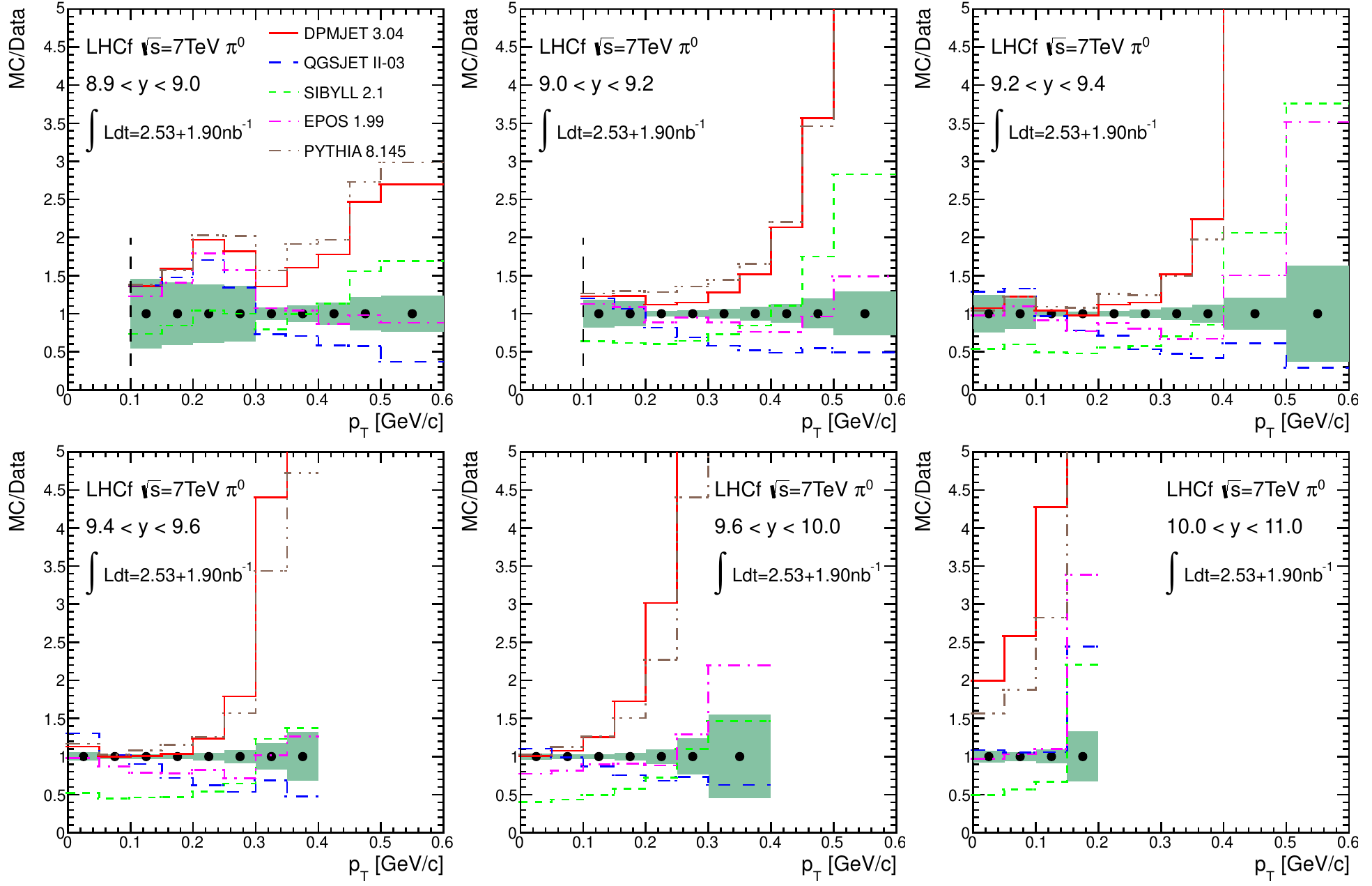}}}
\caption[*]{Ratio of the combined ARM1 and ARM2 $\pi^0$ p$_{\mathrm T}$ spectra to the p$_{\mathrm T}$ spectra predicted by various hadronic interaction models. Shaded areas indicate the range of total uncertainties of the combined spectra. Figure from Ref.~\cite{pizero}.}
\label{fig.pizero}
\end{figure}

The measurement of the single photon energy spectra reported in the previous section are very important to constrain 
the interaction models used in HECR but are not the only important measurement. A deeper insight can be achieved by measuring 
the transverse momentum spectra of photons and $\pi^0$. For this reason, in addition to the measurement of 
the single photon spectra, the LHCf experiment has recently finalised the measurement of the transverse 
momentum spectra for different rapidity bins for $\pi^0$ produced in 7 TeV p-p collisions at LHC. 
The integrated luminosities corresponding to the data used in this analysis are 2.53 nb$^{-1}$ (Arm1) and 
1.90 nb$^{-1}$(Arm2) after the data taking live times were taken into account. The $\pi^0$ are reconstructed in LHCf through the identification of their 
decays in two photons. Events are selected requiring that the two photons enter different calorimeter towers; due to the 
geometrical acceptance of the detector only photons from $\pi^0$ decays with an opening angle of $\theta<0.4$ mrad can be detected. 
Energy, p$_{\mathrm T}$ and rapidity of the $\pi^0$ are reconstructed through the measurement of the photon energy and incident position in 
each calorimeter. In order to ensure good event reconstruction efficiency and geometrical acceptance, the range of the $\pi^0$ 
rapidity and transverse momentum are limited to $8.9<y<11.0$ and $p_{\mathrm T}<0.6$ GeV/c, respectively.
Figure~\ref{fig.pizero} shows the ratios of  p$_{\mathrm T}$ spectra predicted by DPMJET 3.04 (solid, red), 
QGSJET II-03 (dashed, blue), SIBYLL 2.1 (dotted, green), EPOS 1.99 (dashed dotted, magenta), 
and PYTHIA 8.145 (default parameter set, dashed double-dotted, brown) to the combined ARM1 and ARM2 $p_{\mathrm T}$ 
spectra (black dots). Error bars have been taken from the statistical and systematic uncertainties.
Among hadronic interaction models tested in this analysis, EPOS 1.99 shows the best overall agreement with the LHCf 
data, although it behaves softer than the data in the low p$_{\mathrm T}$ region, p$_{\mathrm T} \lesssim 0.4$ GeV/c in $9.0<y<9.4$ 
and  p$_{\mathrm T}\lesssim 0.3$ GeV/c in $9.4<y<9.6$, and behaves harder in the large p$_{\mathrm T}$ region.
DPMJET 3.04 and PYTHIA 8.145 show overall agreement with the LHCf data for $9.2<y<9.6$ and p$_{\mathrm T}<0.25$ GeV/c, 
while the expected $\pi^0$ productions rates by both models exceed the LHCf data for larger p$_{\mathrm T}$. Also 
SIBYLL 2.1 predicts harder pion spectra than the LHCf data, although the expected $\pi^0$ yield is generally small. Finally, 
QGSJET II-03 predicts $\pi^0$ spectra that are softer than the LHCf data and the other models.

By fitting the  p$_{\mathrm T}$ spectra in each rapidity bin it is possible to extract the average transverse momentum,
$<$p$_{\mathrm T}>$, which results to be consistent with typical values for soft QCD processes. 
Comparison between the LHCf and UA7 results indicate an $<$p$_{\mathrm T}>$  versus rapidity that 
is independent of the center of mass energy, in agreement with the expectation of EPOS 1.99, while SYBILL 2.1 tipically gives 
harder $\pi^0$ spectra, namely larger $<$p$_{\mathrm T}>$, and QGSJET II-03 gives softer $\pi^0$ spectra, 
namely smaller $<$p$_{\mathrm T}>$  than the experimental data.

\section{Impact of LHCf results on HECR Physics}
\label{sec:4}

The first LHCf results have raised attention in the HECR community. As reported in 
the previous paragraphs, none of the models agree in the whole energy range with the data. Tuning of the models are 
hence needed to describe the Physics of hadronic interactions at the TeV scale.
In order to better understand the implication of this measurement for the HECR Physics, a collaboration with 
several MC developers and theoreticians has started.

\begin{figure}[htb]
\vfill \begin{minipage}{.49\linewidth}
\begin{center}
\includegraphics[height=.8\textwidth]{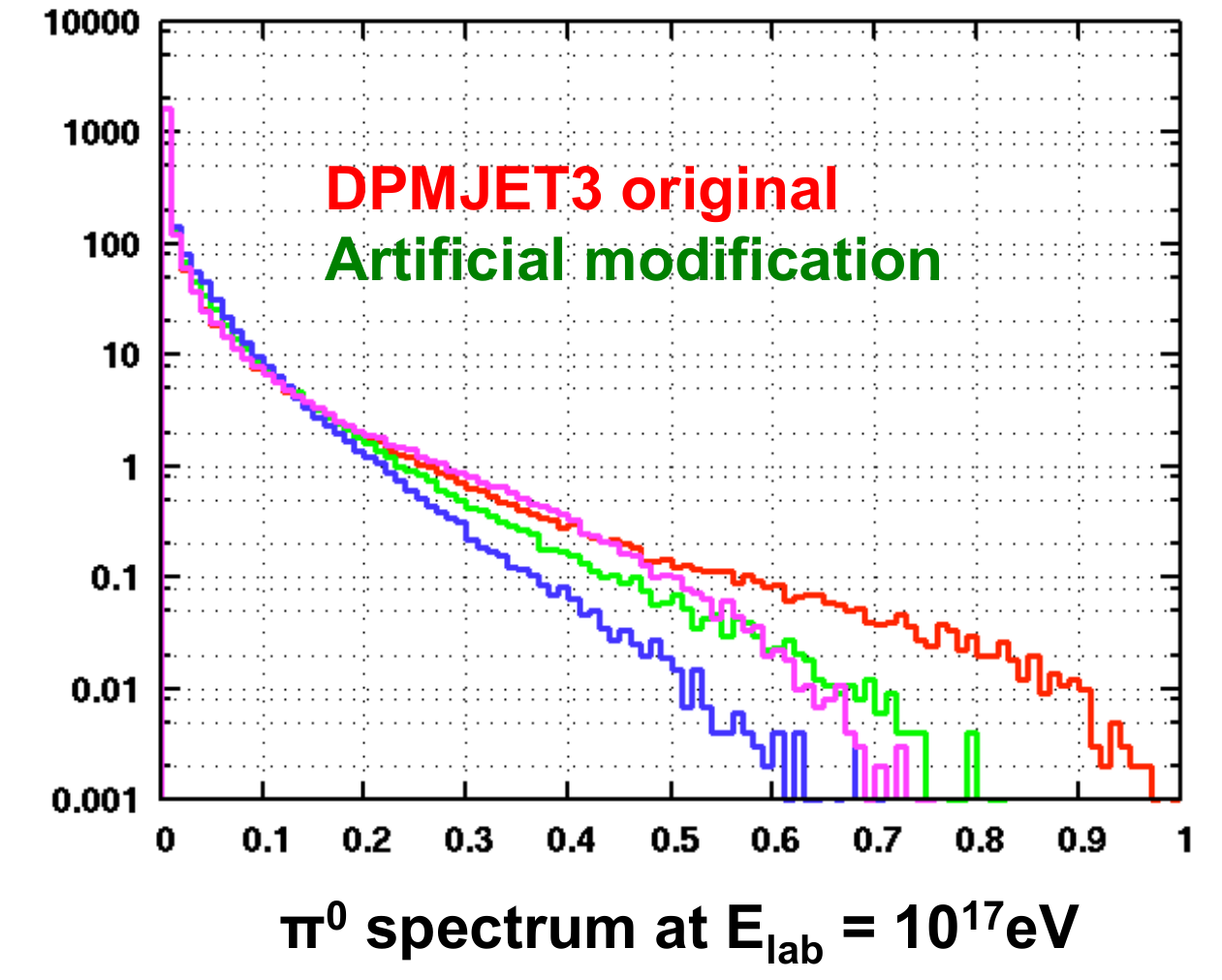}
\end{center}
\end{minipage}\hfill
\begin{minipage}{.49\linewidth}
\begin{center}
\includegraphics[height=.8\textwidth]{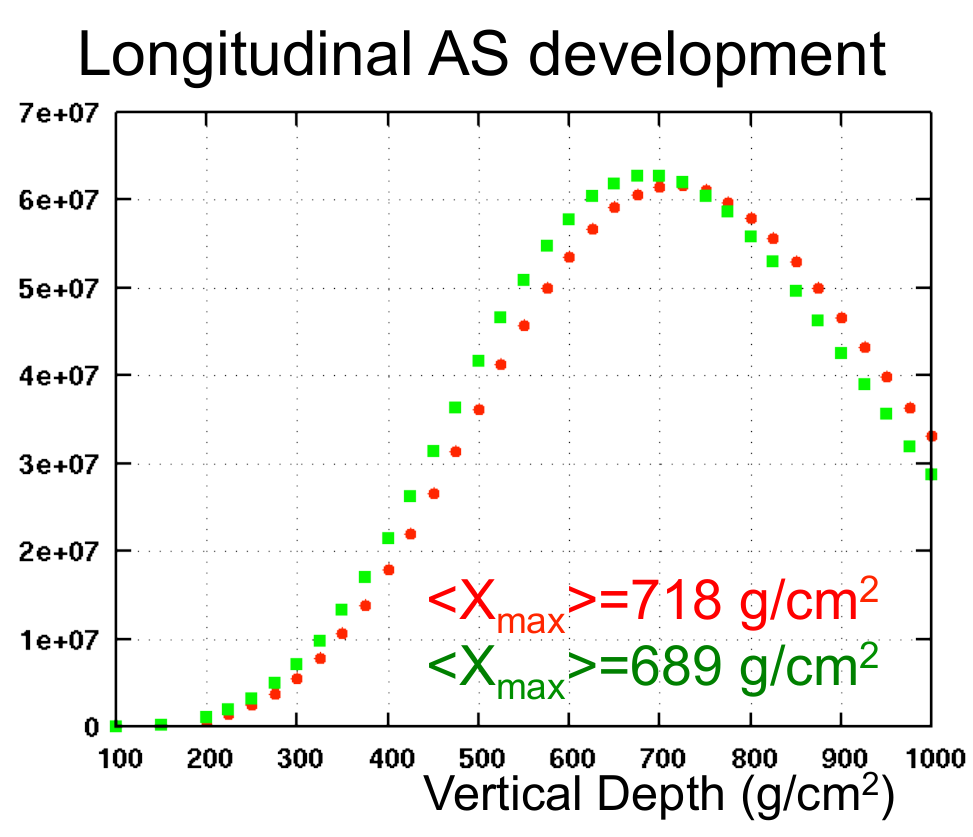}
\end{center}
\end{minipage}
\caption[*]{$\pi^0$ spectra at E$_{lab}=10^{17}$ eV for original DPMJET III-3.04 model (red) and artificially modified models (green, blue, magenta), see text (left). Average longitudinal development of the atmospheric shower in the QGSJET II model (red points) and in the artificially modified model (blue points) (right)}
\label{fig.kasahara}
\end{figure}

As an example, we have artificially modified the DPMJET III~3.04 model to produce a $\pi^0$ spectrum that 
differs from the original one by an amount approximately equal to the difference 
expected between the different models.
Figure~\ref{fig.kasahara} (left panel) shows the $\pi^0$ spectra at E$_{lab}=10^{17}$ eV predicted by the original 
  DPMJET III~3.04 model and the artificially modified ones, while on the right panel  
the average longitudinal development of the atmospheric shower in the original 
model (red points) and in the artificially modified model (green points) is shown.
A difference in the position of the shower maximum of the order of 30 g/cm$^2$ is observed.  

\begin{figure}
\begin{center}
\resizebox{80mm}{!}{\includegraphics{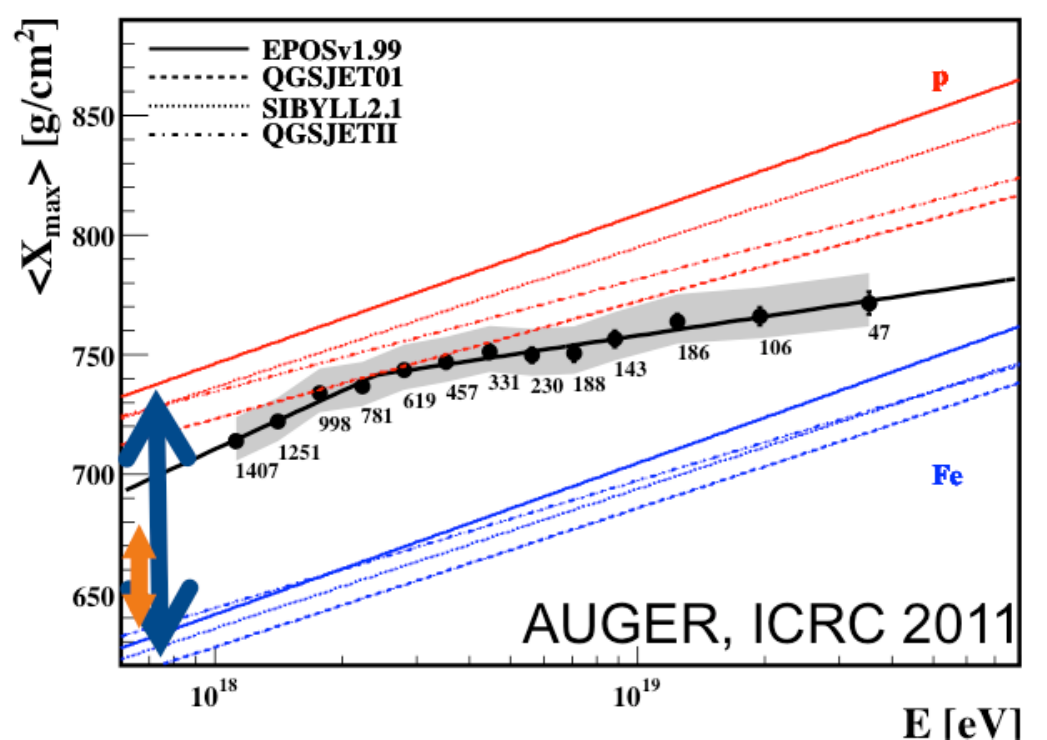}}
\caption[*]{$<{\mathrm X}_{max}>$ distribution as measured by Auger~\protect{\cite{auger}} (black points) 
compared with the model expectations for a light (red) 
or heavy (blue) cosmic ray composition. The yellow arrow correspond to the 30 g/cm$^2$ shift, obtained in Fig.~2.}
\label{fig.composition}
\end{center}
\end{figure}

Fig.~\ref{fig.composition} shows, as an example, the most recent results by Auger Collaboration~\cite{auger} for the distribution of the $<{\mathrm X}_{max}>$ variable as function of the energy, 
which is the most commonly used method to infer cosmic rays composition,  
compared with the model predictions for a proton-like (red lines) and an Iron-like (blue lines) cosmic ray 
components, respectively. The difference in the $<{\mathrm X}_{max}>$ distributions for the two cases is of the order of 
100 g/cm$^2$, hence a 30 g/cm$^2$ shift is a sizable difference which may significantly reflect in the interpretation 
of HECR data. \\
The importance of a direct measurement of the $\gamma$ and $\pi^0$ spectra by LHCf results to be clear. \\

\section{LHCf operation in p-Pb run}
\begin{figure}
\begin{center}
\resizebox{125mm}{!}{\includegraphics{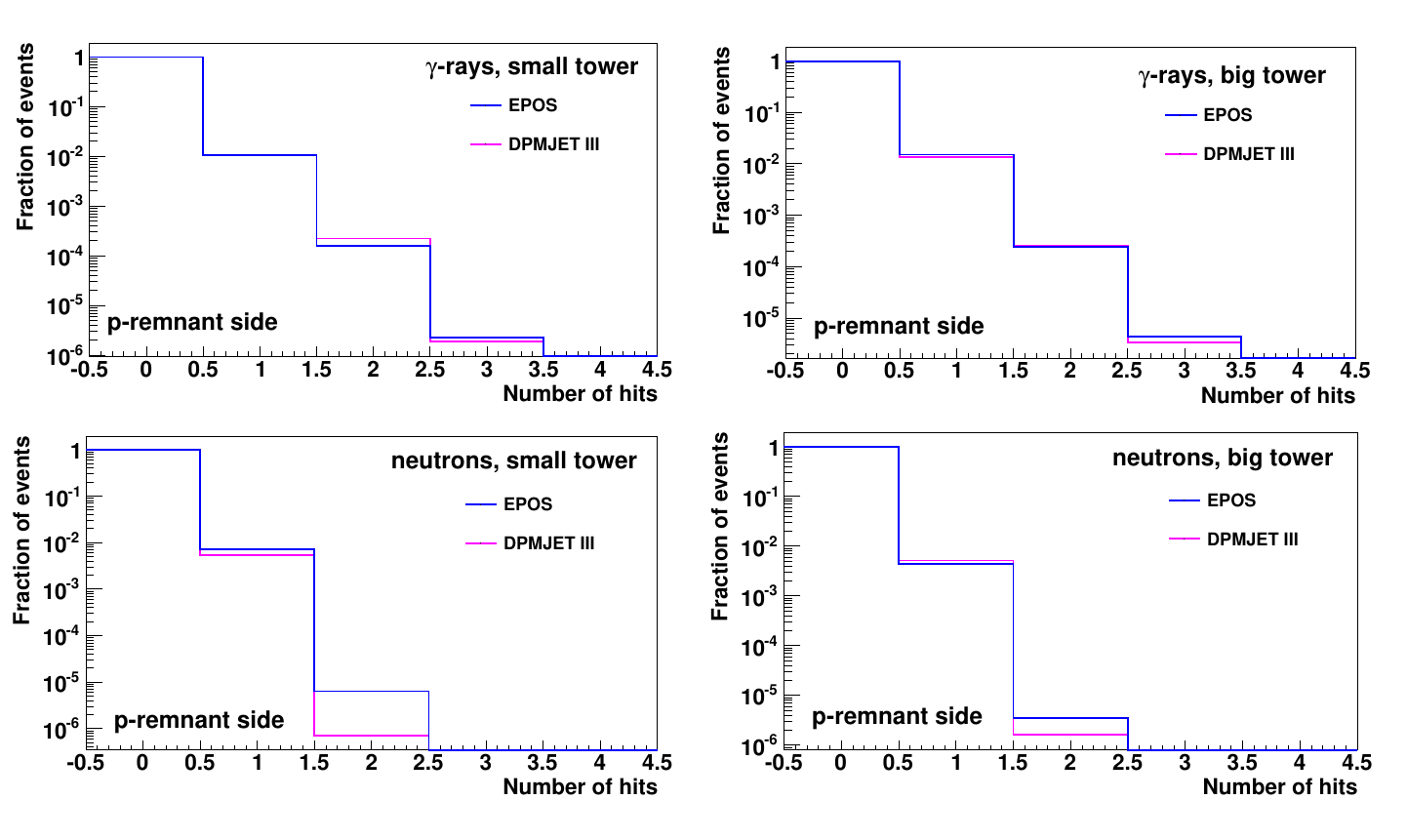}}
\end{center}
\caption[*]{Expected multiplicities on the proton remnant side of photon (top) and neutron (bottom) 
hits on small (left) and large (right) towers in p-Pb collisions at LHC.}
\label{fig.mult_pPb}
\end{figure}

The LHCf detector is also planning to take part to the data taking with p-Pb collisions foreseen at LHC at the beginning of 
2013. The interaction of CRs in the Earth's atmosphere necessarily involves nuclei and not only protons. In fact 
the generation of EAS in the atmosphere is explained in terms of the interaction of primary CRs with atomic nuclei, 
mainly nitrogen and oxygen nuclei, that are present in the atmosphere itself in different percentages. Moreover 
the incoming particles can be simply protons, but can also be light or heavy nuclei. 
The possibility to take data also in pA collisions at high energy at LHC is clearly of extreme importance for 
the understanding of the properties of hadronic interactions and gives to LHCf  the possibility to broaden its Physics program. 
 Despite the ideal probe to study the interaction of CRs with the atmosphere are collisions with nitrogen and oxygen nuclei, 
which unfortunately are not currently available at the TeV energy regime, anyway the study of p-Pb collisions offers a unique 
opportunity to measure the modification induced by nuclear effect in the forward region. \\
The asymmetry of the proton-Lead ion collision leads to critical differences in the events observed by the LHCf detector, 
depending on whether it is installed on the proton-remnant or Lead-remnant side. 
The proton-remnant side is evidently the most interesting from the point of view of CR physics. Fig.~\ref{fig.mult_pPb} shows the 
expected multiplicities of neutral particles hitting the two calorimeter towers of the LHCf Arm2 detector located on the proton-remnant side. Prediction of two models are shown (DPMJET-III and EPOS) for photons (top) and neutrons (bottom) both on the small
tower (left) and the big tower (right).  The expected multiplicities on the proton remnant side are well under control. 
Approximately 1\% of the events have one single hit and less than a fraction 10$^{-5}$ of the events have two hits on a single tower. 

\begin{figure}
\begin{center}
\resizebox{125mm}{!}{\includegraphics{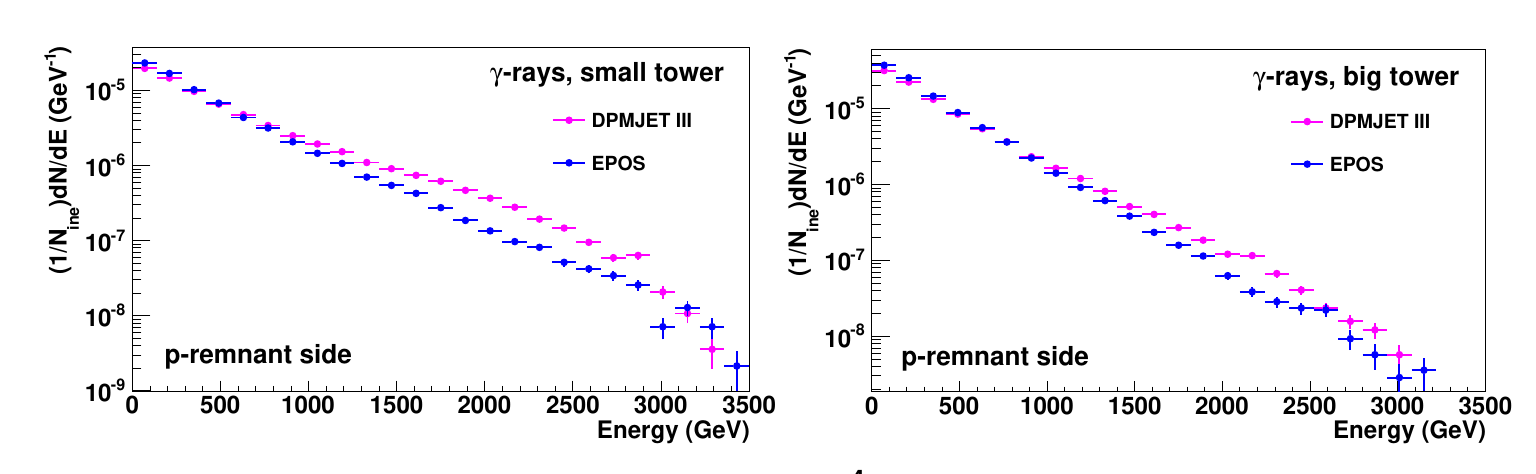}}
\phantom{a}
\resizebox{125mm}{!}{\includegraphics{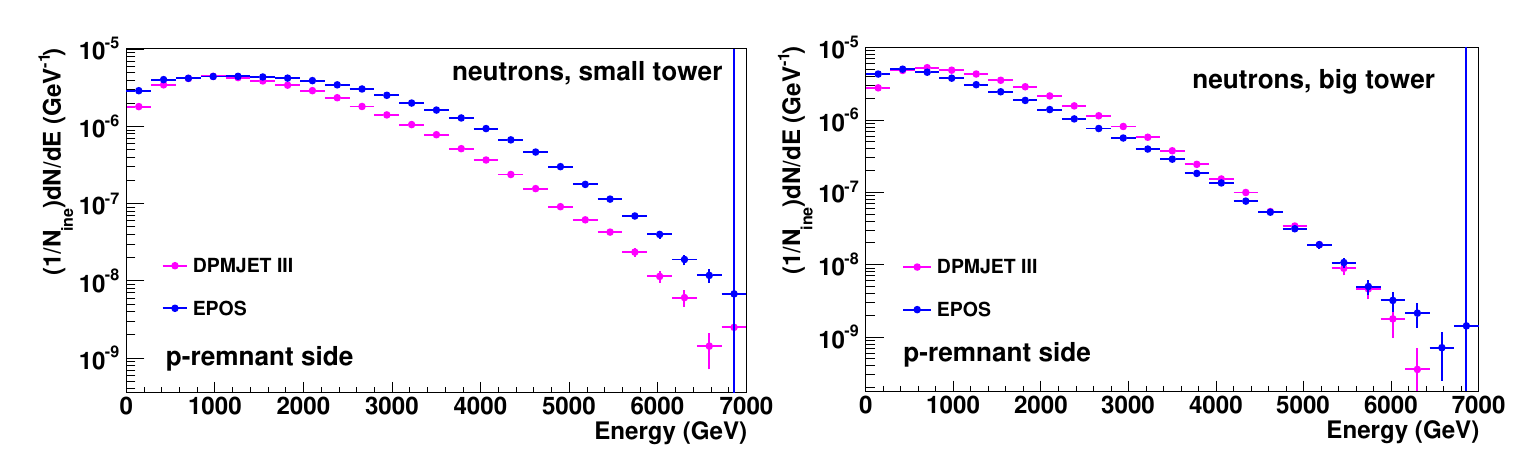}}
\end{center}
\caption[*]{Expected energy spectra of photons (top) and neutrons (bottom) hitting the Arm2 detector on the proton-remnant side 
for the small tower (left) and big tower (right) in the p-Pb collisions. Results of DPMJET-III (magenta) and EPOS (blue) models are shown.}
\label{fig.pPb}
\end{figure}

Fig.~\ref{fig.pPb} shows the expected 
photon (top panel) and neutron spectra (bottom) on the proton remnant side in each of the two calorimeter towers of 
the ARM2 detector for the two different models. The two spectra have been convoluted with the energy resolution of 
the detector. A clear disagreement between the prediction of the two models is visible especially for energies larger than 1 TeV
and the LHCf detector can easily discriminate between them both using photon as well neutron spectra thus providing additional 
useful information for the calibration of such models.

\section{Future activities and summary}
In addition to the analysis and data taking activities, the LHCf Collaboration is working on the upgrade of the detector to improve the 
radiation resistance in view of the 14 TeV p-p run, currently foreseen in 2015. The scintillating part of the 
detector will be replaced with GSO slabs, which has been proved to be more radiation hard~\cite{gso}, 
thus enabling LHCf to sustain the radiation level foreseen in the 14 TeV run.
Additional improvements in the front-end electronics of the silicon position sensitive layers of ARM2 detectors as well 
as an optimization of 
the layout to improve the stand-alone silicon energy resolution are also on going.


\begin{thebibliography}{0}
\bibitem{Takeda} M. Takeda et al., Phys. Rev. Lett. {\bf 81} (1998) 1163.
\bibitem{Abbasi:2007sv} R. U. Abbasi et al., Phys. Rev. Lett. {\bf 92} (2004) 1511.
\bibitem{Abreu:2011pj}  P.~Abreu, et al.  [The Pierre Auger Collaboration], arXiv:1107.4809 [astro-ph.HE].
\bibitem{Tsunesada:2011mp}  Y.~Tsunesada [for the Telescope Array Collaboration],  arXiv:1111.2507 [astro-ph.HE].
\bibitem{UA7} E. Par\'e et al., Phys. Lett. B {\bf 242} (1990) 531.
\bibitem{lhcf.jinst} O. Adriani et al., JINST {\bf 3} (2008) S08006. 
\bibitem{plb} O. Adriani et al., {Phys. Lett.} {\bf B703} (2011) 128.
\bibitem{plb2} O. Adriani et al., {Phys. Lett.} {\bf B715} (2012) 298.
\bibitem{DPMJET} F.W. Bopp et al., {Phys. Rev.} {\bf C77} (2008) 014904.
\bibitem{qgsjet} S. Ostapchenko, {Phys. Rev.} {\bf D83} (2011) 014108.
\bibitem{sybill} E.-J. Ahn et al., {Phys. Rev.} {\bf D80} (2009) 094003.
\bibitem{epos} K. Werner et al., {Nucl.Phys.Proc.Suppl.} {\bf 175-176} (2008) 81.
\bibitem{pythia} T. Sj\"{o}stand, et al., {Comput. Phys. Comm.} {\bf 178} (2008) 852.
\bibitem{pizero} O. Adriani et al., Phys. Rev. D {\bf 86} (2011) 092001.
\bibitem{auger} J.Bellido for the Auger Collaboration, in {\it Proceedings of 32$^{st}$ ICRC, Beijing}, (2011).
\bibitem{LOIpPb} O. Adriani et al.,  CERN-LHCC-2011-015. LHCC-I-021 (2011).
\bibitem{gso} K. Kawade et al.,  JINST {\bf 6} (2011) T09004.
\end{thebibliography}
\end{document}